# TCP Congestion Control Scheme for Wireless Networks based on TCP Reserved Field and SNR Ratio

**Youssef Bassil**

LACSC – Lebanese Association for Computational Sciences, Registered under No. 957, 2011, Beirut, Lebanon

Email: youssef.bassil@lacsc.org

**Abstract** – Currently, TCP is the most popular and widely used network transmission protocol. In actual fact, about 90% of connections on the internet use TCP to communicate. Through several upgrades and improvements, TCP became well optimized for the very reliable wired networks. As a result, TCP considers all packet timeouts in wired networks as due to network congestion and not to bit errors. However, with networking becoming more heterogeneous, providing wired as well as wireless topologies, TCP suffers from performance degradation over error-prone wireless links as it has no mechanism to differentiate error losses from congestion losses. It therefore considers all packet losses as due to congestion and consequently reduces the burst of packet, diminishing at the same time the network throughput. This paper proposes a new TCP congestion control scheme appropriate for wireless as well as wired networks and is capable of distinguishing congestion losses from error losses. The proposed scheme is based on using the reserved field of the TCP header to indicate whether the established connection is over a wired or a wireless link. Additionally, the proposed scheme leverages the SNR ratio to detect the reliability of the link and decide whether to reduce packet burst or retransmit a timed-out packet. Experiments conducted, revealed that the proposed scheme proved to behave correctly in situations where timeouts were due to error and not to congestion. Future work can improve upon the proposed scheme so much so that it can leverage CRC and HEC errors so as to better determine the cause of transmission timeouts in wireless networks.

## 1. Introduction

When the number of packets sent to any network is more than it can handle, congestion develops [1]. Numerous algorithms have been developed since the dawn of computer networking to deal with network congestion. In fact, the simplest way to solve congestion is to employ the principle of packet conversation in which the sender stops sending a new packet until the previous one is successfully delivered to the receiver. For this reason, the TCP protocol, one of the core protocols of the Internet protocol suite, employs the concept of congestion control which dynamically controls the flow of packet inside a network, and prevents network performance collapse [2].

In the early days of data communication, the problem was how to detect congestion because not receiving an acknowledgment from the receiver (referred to as timeout) does not mean that the packet was lost due to congestion but also to noise or error in the transmission wire. However, with the advancement of technologies, packet loss due to transmission error became relatively infrequent as most communication trunks and network infrastructures achieved a high-level of reliability and resilience. As a result, transmission timeouts in modern computer wired networks are 99% due to congestion [3].

The reliability of data transmission has led the TCP protocol to be optimized for wired networks in a sense that any packet loss is considered to be the result of network congestion and not the result of network errors. Algorithmically, when a timeout is detected, the TCP congestion control algorithm reduces dramatically the packet burst to diminish the network load and relieve the congestion. In effect, the TCP congestion algorithm cannot differentiate the loss caused by congestion from the one caused by error. Once a loss occurs, TCP deals with that as a congestion event and halves down the size of its congestion window. This unnecessary slowdown reaction decreases the throughput of TCP and reduces the overall speed of the network.

In practice, the aforementioned behavior of the TCP congestion algorithm is acceptable in wired networks as packet timeouts are most of the time caused by congestion. However, this is totally inappropriate in wireless networks as wireless links are known to experience a lot error bits and packet loss due to fading, interference, hand-off, and other radio effects; consequently, packet loss in wireless networks cannot be considered as due to congestion. As a result, the TCP often makes the wrong decision by slowing down the burst of packets while it should instead retransmit lost packets. This problem is popularly known as the TCP performance problem over wireless network and has been researched and studied by many researchers for many years now [4]. The key in solving this problem is to allow TCP to differentiate between timeouts caused by congestion and those caused by errors and noise in the wireless channel.

This paper proposes a new TCP congestion control scheme suitable for wireless as well as wired networks and is intended to distinguish congestion losses from error losses. The proposed scheme is based on using the reserved bits of the TCP header to indicate whether the established connection is over a wired or a wireless link. Moreover, the scheme harnesses the SNR (Signal-to-Noise) ratio to detect the reliability of the link and decide whether to reduce packet burst or retransmit a timed-out packet. In brief, when a timeout occurs, the TCP protocol checks the type of the link, if it is wired, then packet loss is due to congestion so the classical slow-start congestion control algorithm is executed to decrease the size of the congestion window. On the other hand, if the link is wireless and SNR ratio is less than 5dB, then packet loss is due to error and thus the timed-out packet is retransmitted leaving the congestion window intact. If the link is wireless and SNR ratio is greater than 5dB, then packet loss is due to congestion so the size of the congestion window is reduced to slow down the burst of packets.

## 2. TCP Congestion Control Algorithm

TCP short for Transmission Control Protocol is the most dominant protocol used in computer networking and on the Internet. Characteristically, TCP encompasses various features, one of them is congestion control.

In TCP, when a connection is established between a sender and a receiver, an appropriate window size must be selected. Actually, there exist two windows: the receiver's window and the sender's congestion window. The number of bytes that can be transmitted by the sender is the minimum of these two windows. Hence, in case the receiver's window size is 16 KB, and the sender's congestion window is 8 KB, then the transmission would occur at 8 KB. In contrast, if the receiver's window size is 8 KB and the sender's congestion window is 16 KB, then the transmission would occur at 8 KB. The sender calculates its congestion window size by inspecting the property of the medium network such as delays, traffic, and bandwidth; whereas, the receiver calculates its window size based on its buffer size.

When a TCP connection is established, the sender initializes the congestion window to the size of the maximum segment available on the connection. It then sends one single maximum segment n. If this segment is acknowledged by the receiver before times out, the sender adds another segment doubling the size of its congestion window and sends the two segments 2n. As each of these segments is acknowledged by the receiver, the congestion window is increased by one maximum segment doubling its previous size. Basically, each acknowledged burst doubles the congestion window until either a timeout occurs or the size of the receiver's congestion window is reached. In case of a timeout, the segment size is decreased by half and so for the congestion window. The idea behind this behavior is to maintain, for instance, a window size of 4 KB as long as it is acknowledged by the receiver. Once, it gives a timeout, it is dropped to 2 KB to avoid congestion [5, 6]. Figure 1 is an example of the TCP congestion algorithm.

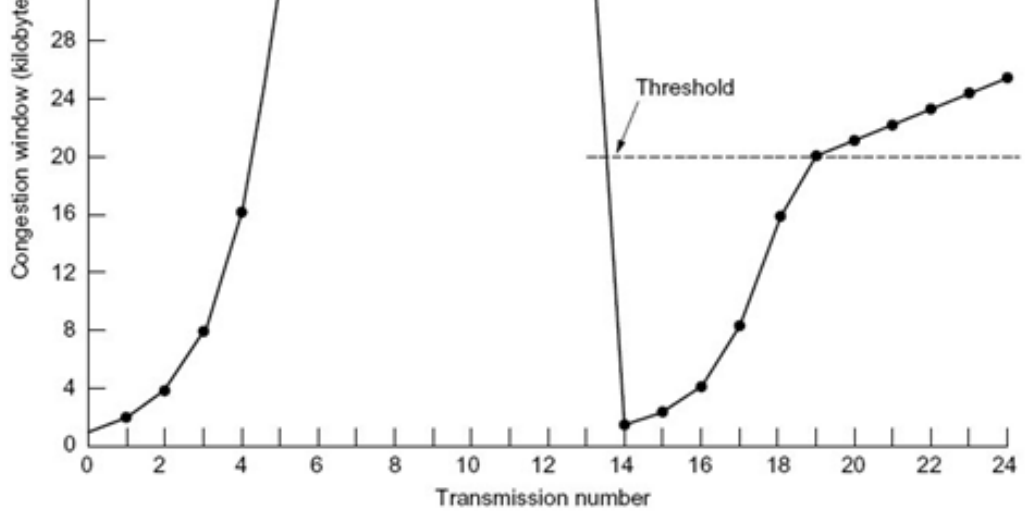

Figure 1. Example of the TCP Congestion Algorithm

## 3. Problem Statement - TCP over Wireless Network

Currently, all implementations of the TCP congestion algorithm assume that timeouts are caused by congestion, and not by transmission errors. This notion stands when applied to wired networks as they are relatively reliable and exhibit very low or no bit errors. However, this notion does not stand for wireless links as they suffer from high error and packet loss rates; and thus, they are still considered significantly unreliable. For this reason, any packet loss in wireless transmission is falsely considered by the TCP protocol as due to congestion which triggers the congestion algorithm to reduce the window size to one segment and consequently reducing transmission speed and packet throughput. For instance, if 25% of all packets are lost, then when the sender transmits 200 packets per second, the throughput is 150 packets per second. If the sender slows down to 100 packets per second, the throughput drops to 75 packets per second [7].

As the TCP congestion algorithm was not initially designed to support the error-prone wireless network, but the very reliable wired network, it is impossible for the sender to differentiate between congestion loss and error loss. As a result, in timeout situations over wireless networks, the TCP often makes the wrong decision by slowing down the burst of packets while it should instead retransmit lost packets.

## 4. Related Work

Several schemes and approaches were extensively studied and experimented to solve the TCP congestion problem over wireless networks. Some of the most successful ones are: I-TCP, Snoop, ECN, WTCP, Westwood, TCP Vegas, TCP Veno, M-TCP, and JTCP.

### *4.1. I-TCP*

The Indirect-TCP (I-TCP) [8] is a modification of the original TCP protocol in which the network connection between sender and receiver is divided into two parts: Wired connection using the standard TCP and wireless connection using a modified version of the TCP. Wired connection is between the fixed host and a middleware station called proxy; while, wireless connection is between the proxy and the mobile hosts. The novelty of this approach is that errors from the wireless connection are corrected at the TCP proxy and do not propagate through the fixed network. The drawback of I-TCP is that it violates TCP's end-to-end semantics and introduces extra overhead as packets are processed twice, one time between the fixed host and the proxy and another time between the proxy and the mobile host. Figure 2 depicts the I-TCP architecture.

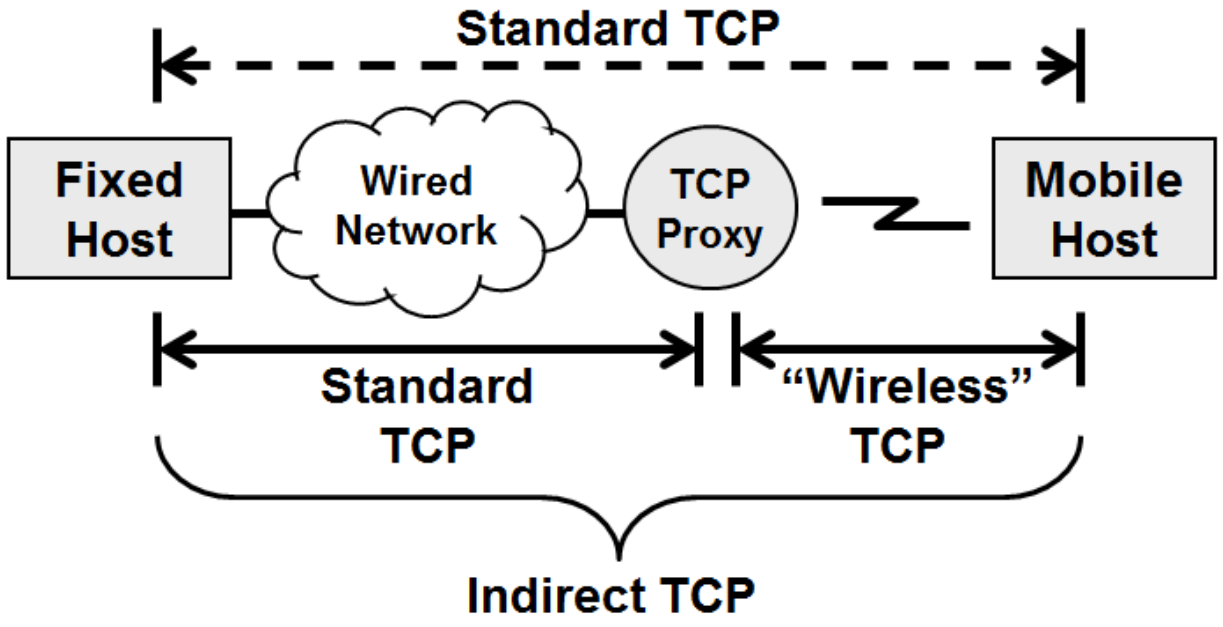

Figure 2. Indirect-TCP Architecture

### 4.2. *Snoop*

Snoop [9] is a classic hiding non-congestion loss method in which a Snoop agent is employed between the wired and the wireless network. When the Snoop agent receives a packet from the wired network, it caches it into an internal buffer and then forwards it to the mobile host. The agent will afterwards wait for acknowledgment from the mobile host. When the agent receives the acknowledgment, it checks the status of the packet, if it is in the cache, then the lost packet is retransmitted to the destination without going back through the wired network; otherwise, the Snoop agent forwards the acknowledgment to the wired source denoting a congestion problem. The main drawbacks of Snoop are that it is not end-to-end semantic and it does not completely isolate wireless link errors from the fixed network.

### 4.3. *M-TCP*

M-TCP (Multicast TCP) [10] is intended for low bit-rate wireless links. In this scheme, every TCP connection is divided in two segments: The first is a TCP connection from the sender fixed host (FH) to the supervisor host (SH) and it uses the standard TCP protocol. The second connection is between the supervisor host (SH) and the mobile host (MH) and it uses an adapted version of TCP. In action, the FH sends a segment to MH, the SH receives it first then forwards it to the MH which acknowledges it upon receipt. Once SH receives the ACK from the receiver, it forwards it to the FH. In case the MH is disconnected, the SH stops receiving the ACK and considers that the MH has been temporarily disconnected and therefore it sends the ACK to the FH containing the window size of 0. The sender then freezes its timers and sends the backed-off persist packet to the SH which in turn responds with a zero window size until it receives some non-zero window size. As soon as the window size gets larger than zero, the SH directly replies with the suitable window size and restarts all its frozen timers. As a result, the sender resumes transmitting at full-speed.

### 4.4. *ECN and Congestion Coherence*

ECN short for Explicit Congestion Notification [11] is a network congestion scheme that utilizes two bits in the IP header and two bits in the TCP header to record the status of the network. In case of congestion, the ECN bits are set to true in the last transmitted packet. When the receiver receives the packet with ECN set to true, it sets the ECN bits of the acknowledgment packet to true and sends it to the sender. The sender then reduces its window size to avoid congestion. Congestion coherence (CC) [12] is an improvement over ECN to differentiate loss. In case the ECN bits are set to true, then the lost packet event is caused by congestion; otherwise, i.e. ECN bits are set to false, the lost packet event is caused by bit error. Although this scheme can determine the cause of loss exactly, it must be implemented inside the TCP protocol as well as network devices such as routers, switches, and stations. Figure 3 shows an ECN example in which several packets are dropped due to congestion. These packets are obviously marked in red denoting that the ECN bits are set to true.

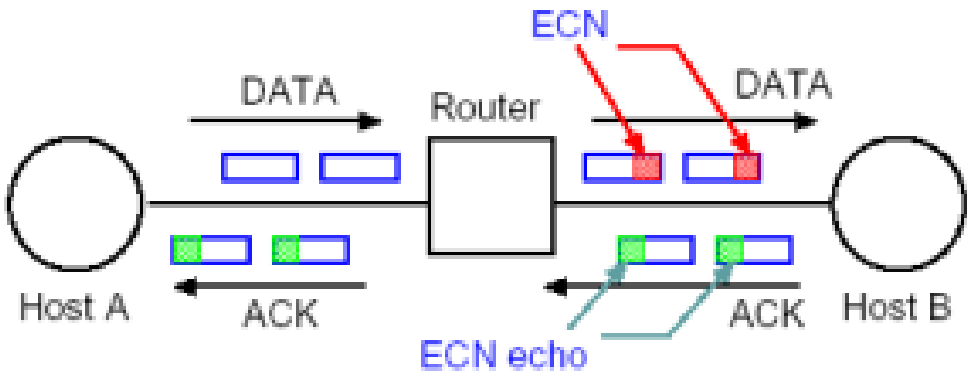

Figure 3. ECN Example

### 4.5. *WTCP*

WTCP (Wireless TCP) [13] is an end-to-end semantic mechanism able to distinguish error losses from congestion losses by comparing the packet arrival time with the packet departure time. WTCP is a replacement for the TCP protocol which, unlike TCP, does not half down its transmission rate for a packet loss but uses inter-packet delay as a metric to compute the rate adaptation at the receiver's end; hence, predicting the cause of packet loss and probing the receiver to find out what packet has to be retransmitted. In that sense, WTCP uses rate-based instead of window-based transmission in which the sender does not decide which packet have to be retransmitted but uses feedback from the receiver to do so. The major drawback of WTCP is that it is completely independent of the original TCP; and thus, it must be implemented by all network nodes and devices.

### 4.6. *Westwood*

TCP Westwood [14] distinguishes the cause of packet loss by guessing the available bottleneck bandwidth. It measures the arrival rate of ACKs and uses the interval of constant feedback ACKs and the packet size to find the optimal network capacity denoted by *SBW[j]*. Additionally, it calculates a smoothed value denoted by *BWE[j]*, by low-pass filtering the sequence of *SBW[j]*.

$$SBW[j] = packet_size \,/\, (current_time - prev_ACK_time)$$
$$BWE[j] = (1 - t) * (SBW[j] + SBW[j-1]) \,/\, 2 + t * BWE[j-1]$$

Where *packet_size* is the size of packet, *current_time* is the most recent time, *prev_ACK_time* is the time of the last ACK, and $t$ is the factor of the low-pass filtering operation. Though Westwood tries to approximate the bandwidth of the connection to best find the congestion window size, it sometimes overestimates the available bandwidth making the overall throughput of the wireless network relatively slow.

### 4.7. *TCP Vegas and TCP Veno*

TCP Vegas [15] estimates the capacity of the TCP link by finding the minimum RTT value called *BaseRTT*. The capacity of the backlog queue is calculated by the following equations:

*Expected = cwnd / BaseRTT*
*Actual = cwnd / ActualRTT*
*Diff = Expected – Actual*
*ActualRTT = BaseRTT + N / Actual*
*N = Actual * (ActualRTT - BaseRTT) = Diff * BaseRTT*

Where *Expected* is the expected rate of a wireless link, *Actual* is the actual rate, *ActualRTT* is the real RTT, *cwnd* is the current TCP congestion window size, and *N* is the size of the queue. The Vegas scheme tries to keep *N* as small as possible by fine-tuning the TCP window size ahead of time; thus, preventing packet loss caused by congestion. On the other hand, TCP Veno [16] improves upon Vegas to judge the cause of loss. It sets a threshold denoted by *thres*, to represent the status of the network. If packet loss occurs and *N>thres*, then the connection is said to be in bad status and the packet will be considered as congestion loss; otherwise, it will be considered as error loss. The disadvantages of TCP Vegas/Veno are that it constantly updates the window size leading to a performance overhead. In addition, it can lead to fluctuation in the window size and round trip times; and consequently, causing delay jitter and inefficient bandwidth utilization.

### 4.8. JTCP

JTCP [17] short for Jitter TCP is a TCP congestion scheme used to detect packet loss and identify whether they are caused by congestion or bit error. It is based on the jitter ratio and packet-by-packet delay that are determined by the inter-arrival jitter i.e. the packet spacing at the sender compared with the packet spacing at the receiver for a pair of packets. The inter-arrival jitter can be calculated as follows:

$$D(i,j) = (R_j - R_i) - (S_j - S_i) = (R_j - S_j) - (R_i - S_i)$$

Where $i$ and $j$ denote the index of packet, $S_i$ denotes the sending time for packet $i$, and $R_i$ denotes the receiving time for packet $i$. JTCP can infer the packet with longer transmitted time and delays it into the router queue until congestion is resolved. JTCP has better performance than other congestion mechanisms; however, it is required to be implemented in the transport portion of the TCP stack at every endpoint, in addition to network devices, stations, and nodes.

## 5. Proposed Solution

This paper proposes a TCP congestion control scheme suitable for wireless as well as wired network environments. It is based on using one single bit of the reserved bits of the TCP header to indicate the type of the link over which a connection is established. If the link is wired, the TCP reserved bit is set to 0 denoting a wired mode; whereas, if the link is wireless, the bit is set to 1 denoting a wireless mode. Additionally, the scheme uses the SNR (Signal-to-Noise) ratio to detect the reliability of the link. In wired mode, any timeout is considered a congestion loss; and thus, congestion is avoided by using the classical TCP start-slow algorithm. However, in wireless mode, two scenarios are possible for a packet timeout, both of which are based on SNR ratio: In case SNR is high, i.e., greater than 5dB, that means that the link is reliable and the loss is due to congestion, so the classical TCP congestion mechanism is executed to slow down the burst of packets. In case SNR is low, i.e., less than 5dB, that means that the link is unreliable and the loss is due to error, so the timed-out packet is retransmitted by the sender.

### 5.1. The Reserved Bits

Transmission Control Protocol or TCP for short is one of the core protocols in the internet protocol suite. Its major role is to provide reliable and ordered delivery of data stream from an application on one computer to another application on another computer [18]. TCP is part of the transport layer which receives a data stream from the application layer for processing. The stream is segmented into small portions called packets, and appended to a TCP header creating a TCP segment. A TCP segment is made out of a header and a data section. The header section consists of 10 compulsory fields, and an optional extension field; while, the data section trails the header and carries the payload data for the application.

The TCP header contains several fields; the source port, destination port, sequence number, data offset, flags, and checksum are few to mention. One overlooked field is the reserved field which is composed of three bits that are reserved by the original TCP design for future use. In modern TCP implementations, these bits are never used and should be set to zero. The proposed scheme exploits this reserved field to detect the type of the link over which the connection is established. In fact, only the first bit is used which can be either set to 0 to indicate a wired communication or to 1 to indicate a wireless communication. This mechanism must be implemented in the TCP stack of every network device including operating system, switches, routers, and stations so that packets are marked as “wired” or “wireless”. Figure 4 depicts the original TCP header format along with the reserved bits leveraged by the proposed TCP congestion scheme.

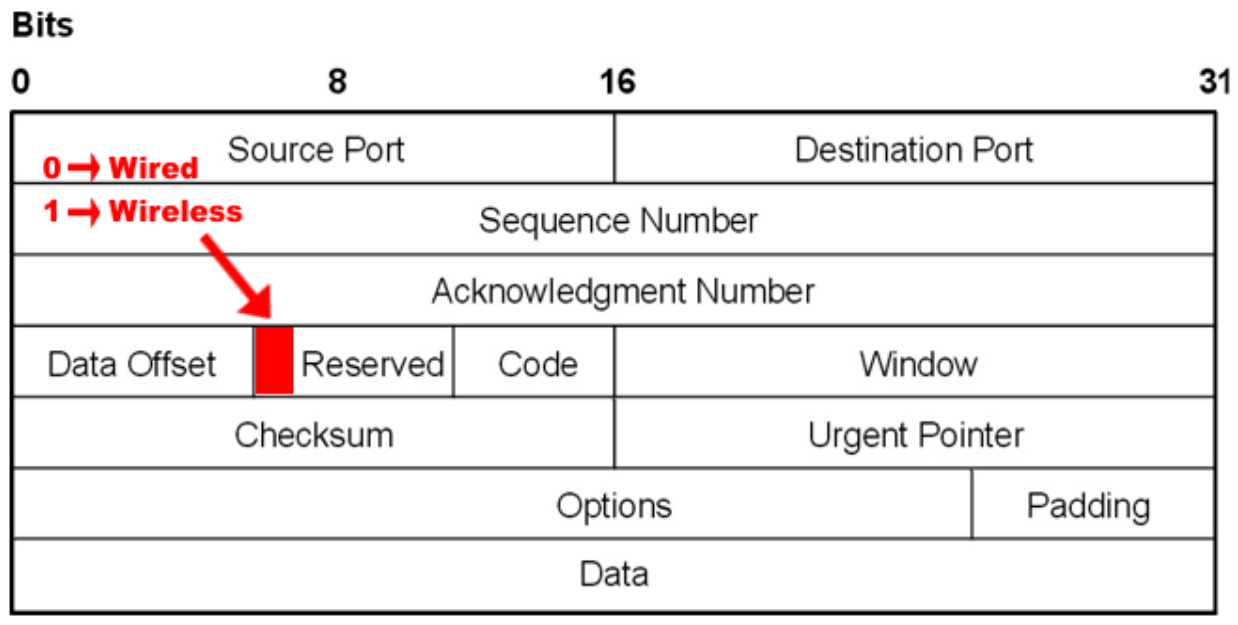

Figure 4. TCP Header with Reserved Bits

### 5.2. The Congestion Algorithm

In digital communication, a signal-to-noise ratio (often abbreviated as SNR) is a measure used to compare the level of a desired signal to the level of background noise [19]. Higher numbers are better. The lower limit of SNR is usually considered to be around 5dB, while 12dB is fairly enough for most conditions. In practice, the SNR ratio is detected upon the initial communication session setup phase. Generally, low SNR leads to high bit and CRC error rate causing packet loss and timeouts. Mathematically, SNR is defined as the ratio of signal power to the noise power, and it can be calculated using the equation below.

$$\mathrm{SNR} = \frac{P_{\mathrm{signal}}}{P_{\mathrm{noise}}}$$

Often SNR is expressed in decibel units according to the following equation:

$$\mathrm{SNR_{dB}} = 10\log_{10}\left(\frac{P_{\mathrm{signal}}}{P_{\mathrm{noise}}}\right) = P_{\mathrm{signal,dB}} - P_{\mathrm{noise,dB}}$$

The proposed scheme when operating in wireless mode exploits the SNR ratio of the communication line to decide whether a timed-out packet was due to congestion or error loss. When a TCP connection first begins, the congestion algorithm initializes the congestion window *cwnd* to one segment. Then, the first bit of the reserved field is set according to the type of the link, i.e., *b=0* for wired and *b=1* for wireless connection. Packets are then sent to the receiver. When they are successfully acknowledged, the congestion window *cwnd* is incremented by one segment, making its size two segments, then four segments, then eight segments, and so on doubling each time the window size until the size advertised by the receiver is reached or until congestion occurs. When congestion occurs, packets will start to be lost triggering a timeout at the sender. In this situation, the sender immediately checks the reserved bit *b*. If it is equal to 0 (wired link), then timeout is considered to be due to congestion. The size of the congestion window *cwnd* would be set to one half of the current size and the sender resumes the burst of packets. In contrast, if the reserved bit *b* is equal to 1 (wireless link), the SNR ratio of the connection is checked. In case it is within a high range, i.e., greater than 5dB (SNR>5dB), then timeout is considered to be due to congestion and the window size is halved by one segment and next packet is sent. However, if SNR ratio is within a low range, i.e., less than 5dB (SNR<5dB), then timeout is considered to be due to error and the timed-out packet is retransmitted to the receiver. The flowchart of the proposed congestion algorithm is depicted in Figure 5; while, its pseudo-code is outlined as follows:

1. Initialize congestion window *cwnd* to one segment.
2. Start sending packets to receiver.
3. For each acknowledgement received, increment congestion window *cwnd* by one segment.
4. If a timeout occurs for a particular packet *p* then check the value of the reserved bit *b* for packet *p*.
    1. If *b=0 (wired link)*, then set *cwnd* to half of its previous size (indicating a congestion)
    2. Else if *b=1 (wireless link)*, then check the SNR ratio of the link.
        1. If SNR>5dB, then set *cwnd* to half of its previous size (indicating a congestion)
        2. Else if SNR<5dB, then don't change the current size of *cwnd*, and retransmit packet *p* (indicating an error)

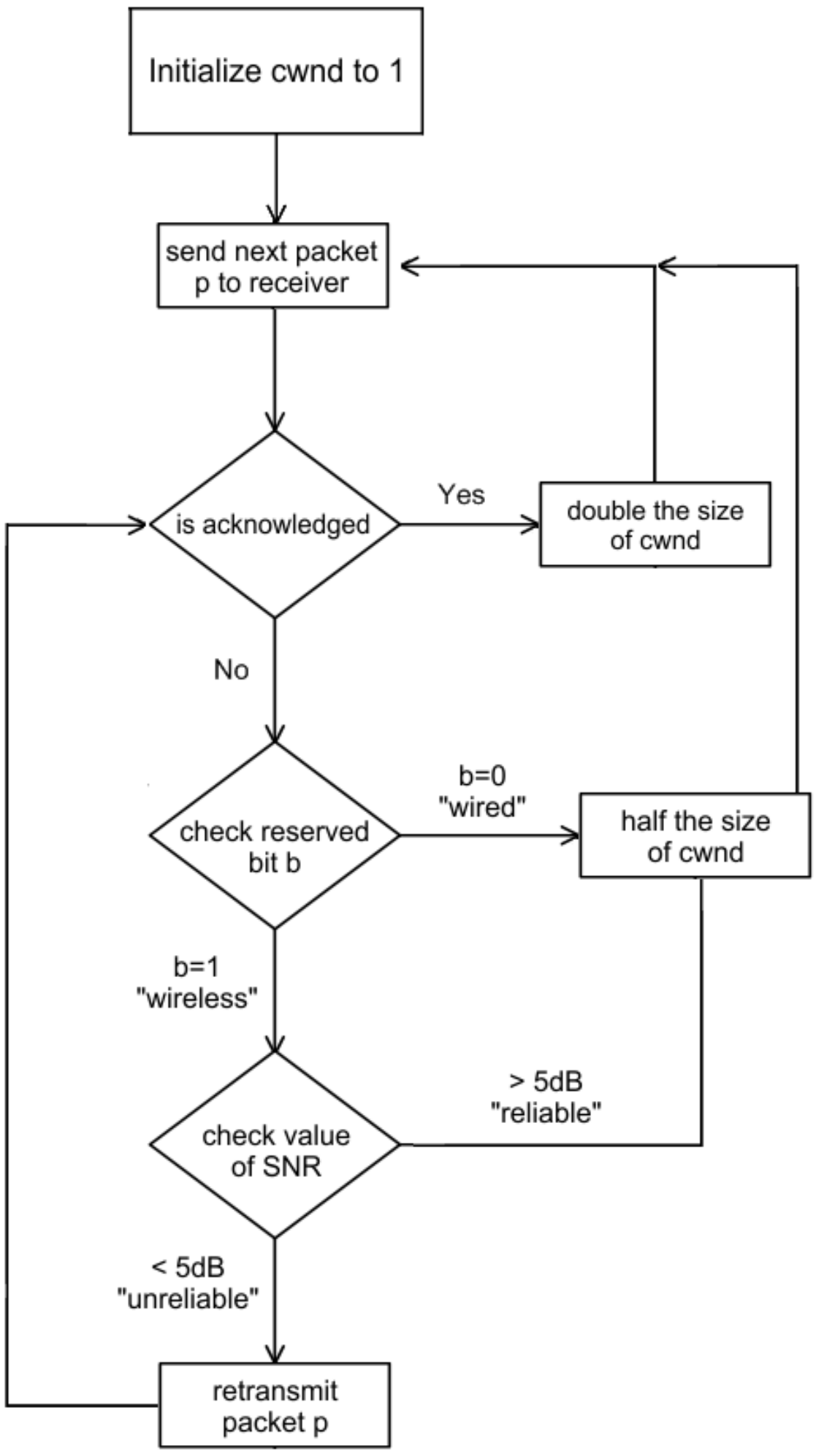

Figure 5. Flowchart of the Proposed Congestion Algorithm

### 5.3. *Advantages*

End-to-end semantics: The proposed scheme maintains true end-to-end semantics, without involving any intermediary between the sender and the receiver.

Performance: The proposed scheme maintains a high level of performance especially that no intermediate nodes are involved which eliminates any extra processing overhead and the need for extra buffer space.

No change in code: The proposed scheme maintains the source code of the sender and the receiver without the need to alter and recompile the communicating applications.

Compatibility: The proposed scheme maintains the original TCP structure with one exception is the integration of the proposed congestion algorithm into the standard TCP RFC 793 stack so as to take advantage of the TCP reserved field and the SNR ratio of the underlying transmission link.

## 6. Simulation & Results

As a proof of concept, the proposed scheme and the standard TCP (RFC 793) [20] were simulated using Network Simulator 3 (NS-3) [21]. For this purpose, a network model was built. It is a wireless topology network in which *C* is a client node, *R1* is a router, and *R* is a receiver node. The bandwidth between *C* and *R1* is 100Mbps with 5ms propagation time; while, bandwidth between *R1* and *R* is 80Mbps with 2ms propagation time. The packet size is 1 KB and the size of queue is 100. Figure 6 depicts the wireless network model to be simulated.

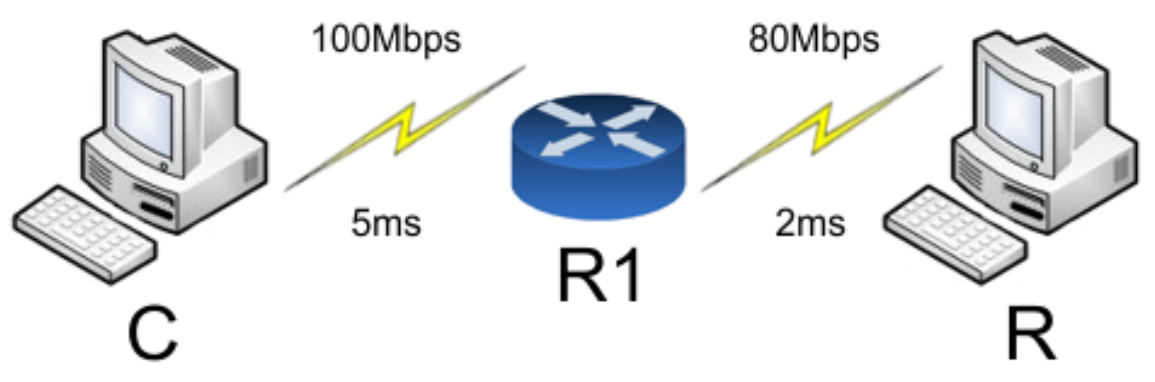

Figure 6. Wireless Network Model

The simulation was run for 350 seconds and two aspects were evaluated in presence of packet loss caused by error: The network throughput and the congestion window size. Figure 7 shows the simulation results for network throughput of the proposed scheme and the standard TCP protocol (RFC 793).

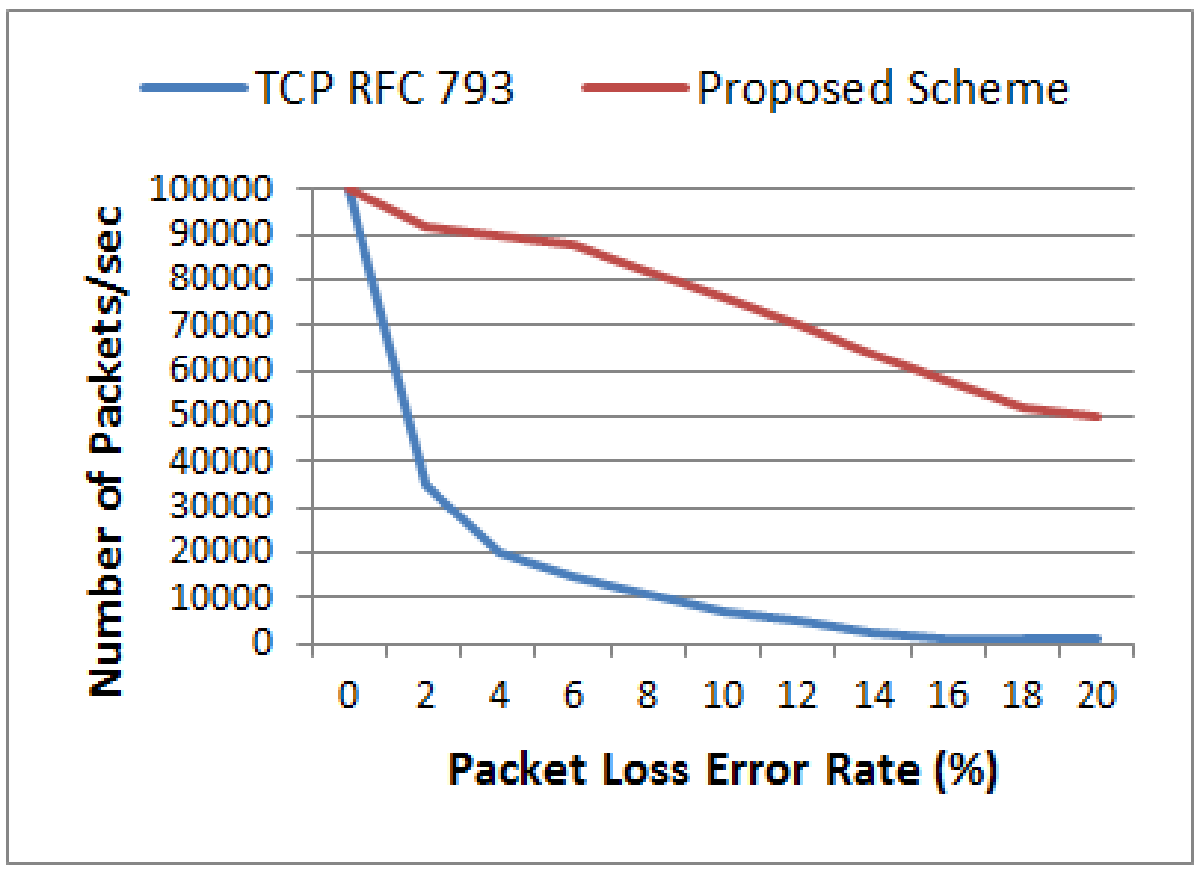

Figure 7. Simulation Results for Network Throughput

In the above results, the throughputs of both schemes are about the same for 0% packet loss rate. However, when the packet loss rate increased, the deviation became more visible. In fact, the throughput of the TCP RFC 793 dropped exponentially with the increase of packet loss; while, the proposed scheme retained a more stable throughput throughout the simulation. This can be explained by the fact that TCP RFC 793 halves down its window size each time a timeout occurs (unacknowledged packet due to congestion loss), leading to a decrease in network throughout. On the other hand, the proposed scheme can detect the rate of packet loss by finding the SNR ratio of the link and consequently decide whether to retransmit the packet or halve down the connection bandwidth. In the simulation, the packet loss was due to error and not to congestion. The high number of packet loss means a low SNR ratio. Therefore, the proposed scheme retransmitted timed-out packets rather than decreasing the window size and the connection bandwidth.

Furthermore, the simulation was run for another 350 seconds to evaluate the congestion window size in presence of packet loss. Figure 8 and 9 show the simulation results for congestion window size of the proposed scheme and the standard TCP protocol (RFC 793).

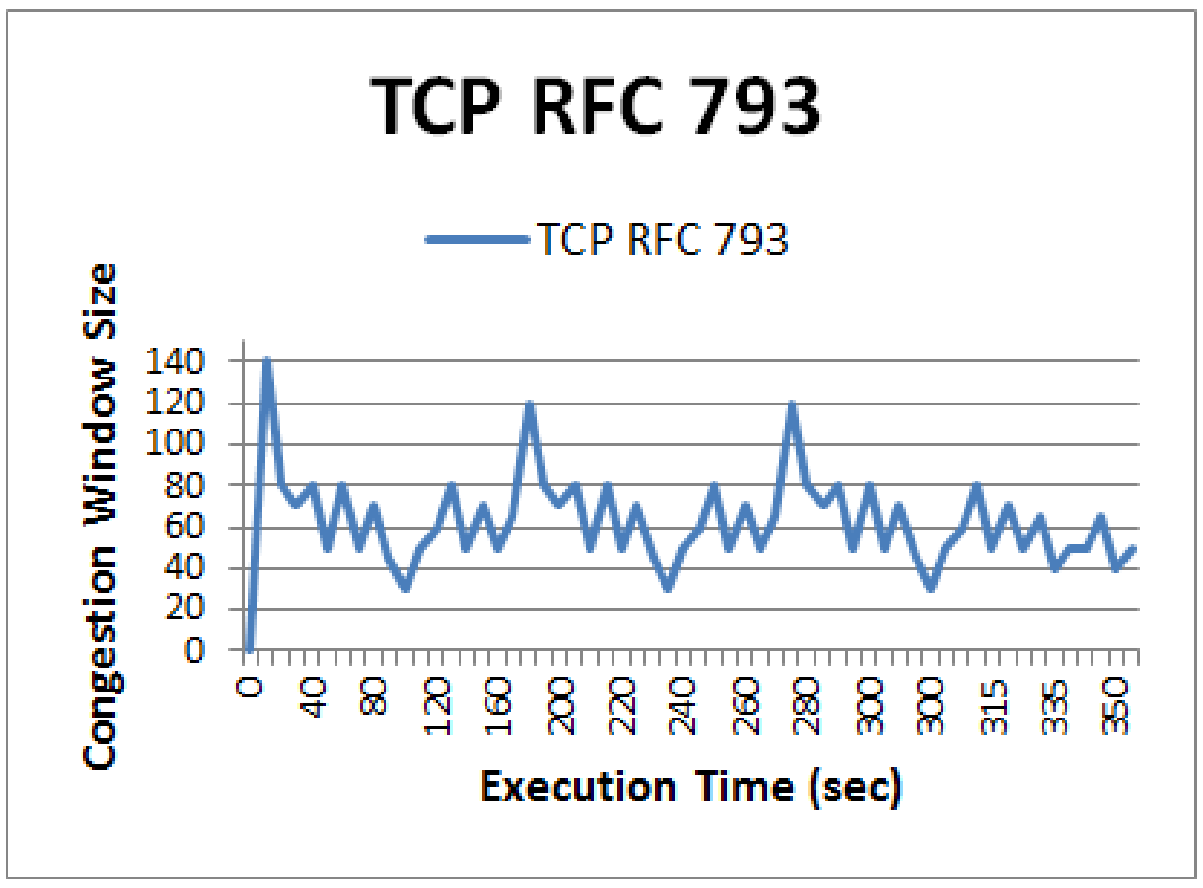

Figure 8. Simulation Results for Window Size of TCP RFC 793

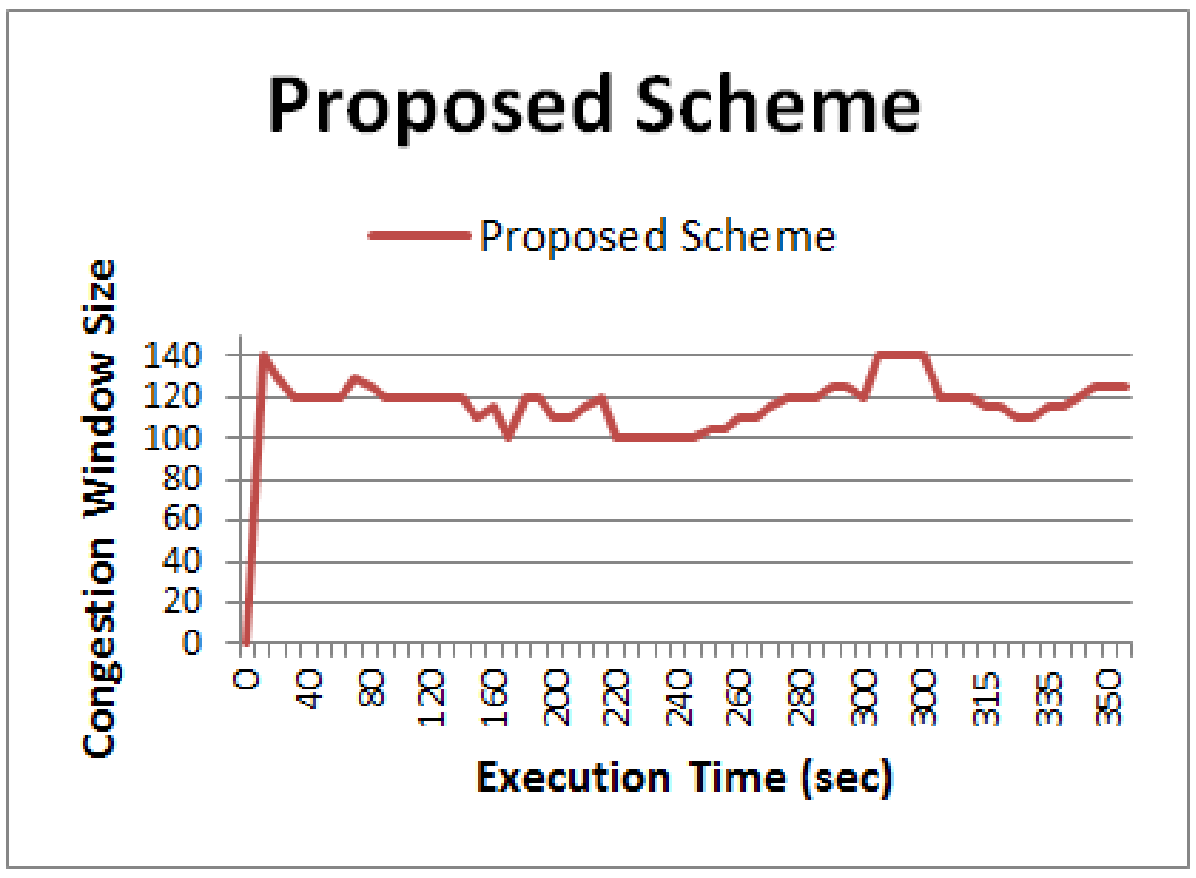

Figure 9. Simulation Results for Window Size of the Proposed Scheme

In the above results, as with the increase of execution time, the packet loss increased which led the TCP RFC 793 to consider the situation as congestion and started reducing the size of the congestion window. It first started at 140 bytes then it went down to somewhat close to 20 bytes. The consequences of this reduction are a lower throughput and a decrease in the connection speed. In contrast, as the proposed scheme can differentiate between packet loss due to congestion and the one due to error, the window size was maintained most of the time with little fluctuation that ranged between 140 and 100 bytes. This fluctuation in the window size was due to timed-out packets in presence of a high SNR ratio (SNR>5dB), indicating no error but network congestion. Nevertheless, the window size was maintained in the presence of timeouts that were due to low SNR ratio (SNR<5dB), indicating no congestion but network errors.

## 7. Conclusions & Future Work

This paper presented a novel scheme for solving the TCP performance problem over wireless networks. Its aim is to allow the TCP protocol to distinguish between transmission timeouts due to congestion and those due to error. The scheme uses the TCP reserved field to identify the type of network over which communication is occurring, in addition to the SNR ratio to determine the reliability of the link so that better decision can be made regarding whether to reduce packet burst or retransmit a timed-out packet. Simulation of the proposed scheme clearly showed that it successfully managed to determine the cause of packet loss in wireless networks and take the right decision in situations where timeouts were due to error and not to congestion, that is, retransmitting the timed-out packet instead of reducing the congestion window size and consequently not diminishing the burst of packets and the overall throughput of the network.

As future work, CRC and HEC metrics are to be added as additional parameters to the proposed scheme so that it can better determine and predict the cause of transmission timeouts in wireless networks.

## Acknowledgment

This research was funded by the Lebanese Association for Computational Sciences (LACSC), Beirut, Lebanon, under the “TCP for Wireless Networks Research Project – TWNRP2012”.

## References

[1] R. Stewart, C. Metz, “SCTP: new transport protocol for TCP/IP”, IEEE Internet Computing, vol. 5, no. 6, pp. 64-69, 2001.
[2] Van Jacobson, and Michael J. Karels, “Congestion Avoidance and Control”, Proceedings of ACM SIGCOMM '88, pp: 314-329, 1988.
[3] Larry L. Peterson, Bruce S. Davie, Computer Networks, A Systems Approach, 5th ed, Morgan Kaufmann, 2011.
[4] G. Xylomenos, G.C. Polyzos, P. Mahonen, and M. Saaranen, “TCP Performance Issues over Wireless Links”, IEEE Communications Magazine, 2001.
[5] M. Allman, V. Paxson, and W. Stevens, “TCP Congestion Control”, RFC 2581, 1999.
[6] Tanenbaum, Wetherall, Computer Networks, 5th ed, Prentice Hall, 2010.
[7] Ye Tian, Kai Xu, and Ansari N, “TCP in Wireless Environments: Problems and Solutions”, IEEE Communications Magazine, vol. 43, no. 3, pp. 27-32, 2005.
[8] Bakre, Badrinath, “I-TCP: Indirect TCP for Mobile Hosts”, In Proceedings of ICDCS 95, 1995.
[9] Hari Balakrishnan, Srinivasan Seshan, Elan Amir, and Randy H. Katz, “Improving TCP/IP Performance over Wireless Networks”, Proceedings of the 1st ACM Conference on Mobile Computing and Networking, Berkeley, CA, November 1995.
[10] K. Brown and S. Singh, “M-TCP: TCP for Mobile Cellular Networks”, ACM Computer Communications Review, vol. 27, no. 5, pp. 19 43, 1997.
[11] Ramakrishnan and Floyd, “A Proposal to add Explicit Congestion Notification (ECN) to IP”, Internet Draft, January 1999.
[12] Chunlei Liu, and Jain R., “Approaches of Wireless TCP Enhancement and A New Proposal Based on Congestion Coherence”, System Sciences, 2003.
[13] P. Sinha, N. Venkitaraman, R. Sivakumar, and V. Bharghavan, ‘‘WTCP: A Reliable Transport Protocol for Wireless Wide-Area Networks’’, ACM Mobicom, Seattle, WA, 1999.
[14] Saverio Mascolo, Claudio Casetti, Mario Gerla, M. Y. Sanadidi, and Ren Wang, “TCP Westwood: Bandwidth Extimation for Enhanced Transport over Wireless Links”, ACM Mobicom, 2001.
[15] Lawrence S. Brakmo, and Larry L. Peterson., “TCP Vegas: End to End Congestion Avoidance on a Global Internet”, IEEE Journal on selected areas in communications, vol. 13, no. 8, 1995.
[16] Cheng Peng Fu, and Soung C. Liew, “TCP Veno: TCP Enhancement for Transmission over Wireless Access Networks”, IEEE Journal on selected areas in communications, vol. 21, no. 2, 2003.
[17] Wu E.H.-K., and Mei-Zhen Chen, “JTCP: Jitter-based TCP for Heterogeneous Wireless Networks”, Selected Areas in Communications, IEEE Journal, vol. 22, no. 4, pp. 757-766, 2004.
[18] G. R. Wright, and W. R. Stevens, TCP/IP Illustrated, Volume 2 (The Implementation), Addison Wesley, 1995.
[19] Rafael C. González, Richard Eugene Woods, Digital Image Processing, Prentice Hall, 2008.
[20] J. Postel, Transmission Control Protocol, RFC 793, September 1981.
[21] NS-3 network simulator, URL: http://www.nsnam.org/